\documentclass[12pt]{article}
\usepackage{epsfig,picinpar,floatflt,amssymb}
\def\bd{\begin{displaystyle}}
\def\ed{\end{displaystyle}}
\def\ba{\begin{array}}

\def\ea{\end{array}}
\def\EQ{\begin{equation}}
\def\EN{\end{equation}}
\def\bea{\begin{eqnarray}}
\def\eea{\end{eqnarray}}
\def\beano{\begin{eqnarray*}}
\def\eeano{\end{eqnarray*}}

\begin{document}
\oddsidemargin 5mm
\newpage     
\pagestyle{empty}
\begin{titlepage}
\begin{flushright}
ISAS/EP/97/108 \\
IC/97/149
\end{flushright}
\vspace{0.5cm}
\begin{center}
{\large {\bf Non--perturbative Results on Universal Quantities \\
              of Statistical Mechanics Models}}\footnote{Invited 
talk given at the XII International Congress of
Mathematical Physics, 13-19 July 1997, Brisbane, Australia. 
Work done under partial support of the EC TMR
Programme {\em Integrability, non--perturbative effects and symmetry 
in Quantum Field Theories}, grant FMRX-CT96-0012} \\
\vspace{1.5cm}
{\bf G. Mussardo$^{a,b,c}$} \\
\vspace{0.8cm}
$^a${\em International School for Advanced Studies, Via Beirut 2-4, 
34013 Trieste, Italy} \\ 
$^b${\em Istituto Nazionale di Fisica Nucleare, Sezione di Trieste}\\
$^c${\em International Centre of Theoretical Physics \\
Strada Costiera 12, 34014 Trieste }\\
\end{center}
\vspace{6mm}
\begin{abstract}
\noindent
Exact calculations of some universal quantities of two--dimensional 
statistical models in the vicinity of their fixed points are illustrated. 
\end{abstract}
\vspace{5mm}
\end{titlepage}
\newpage

\setcounter{footnote}{0}
\renewcommand{\thefootnote}{\arabic{footnote}}

\vspace{1cm}

One of the most important successes of Quantum Field Theory (QFT) 
in recent years consists of the quantitative analysis of the universality
classes of two--dimensional statistical mechanics models close to their 
second order phase transition points. Right at the critical points, 
where the correlation length $\xi$ diverges, methods of Conformal 
Field Theories (CFT) are particularly powerful to determine exactly   
the spectrum of anomalous dimensions, structure constants of 
the OPE algebra, correlation functions etc. i.e. all relevant 
universal quantities which characterise the strong interactions 
of the order parameter fluctuations \cite{BPZ}. The hypothesis of 
conformal invariance led to the explicit solution of a large number 
of important systems in statistical mechanics as conformally invariant 
field theories, among which the Ising model, the Ising model with 
annealed vacancies at its tricritical point, models with $S_q$ and 
$O(n)$ ($n\leq 2$) symmetries, particular limits thereof 
(percolation and Self--Avoiding Walks), Yang--Lee Edge
Singularity, Ashkin--Teller model, RSOS and WZW models, to name a few
(see, for instance \cite{DiFrancesco}). 

For many scopes, however, the data provided by CFT do not exhaust 
all physical information relative to the phase transitions. In a real 
sample, conformal invariance may be broken by the presence of 
impurities, by finite--size effects or, simply, by an imperfect 
fine--tuning of the experimental knobs. It is in any case necessary 
to slightly move the systems away from criticality in order to study their 
responses to external fields. These responses are generally 
encoded in a set of different susceptibilities and scaling functions 
of the statistical models. On a more theoretical level, a perturbation 
of the conformal action is required to investigate the space of the 
coupling constants and its topology consisting of the location of the 
fixed points and the Renormalization Group flows which connect them. 
As we will briefly illustrate in the following, the study of the 
statistical models away from criticality is a very rich subject, on 
which a lot of progress has been recently achieved. Since the breaking 
of scale invariance is associated to the appeareance of some mass scales 
into the problem, a preliminary important question regards the degree 
of universality exhibited by the off--critical systems. Although this 
quality is in general spoiled --- each model presenting in fact some 
sensitivity to the specific lattice realization, to the presence of 
sub--leading interactions, etc. ---  a certain universal behaviour 
may be however observed if one keeps sufficiently close to the critical 
point, in the so--called {\em scaling region} where the correlation 
length $\xi$ is finite but large enough to still discard  
the microscopic details of the models. Hence, as far as the attention 
is restricted to the scaling region, QFT can be safely applied to 
characterise the off--critical massive exitations and their dynamics. 
For the very nature of its methods, QFT should provide one 
of the most efficient ways to compute {\em universal ratios}, i.e. 
pure numbers characterising the scaling region of each universality class. 
Some examples will do more than an abstract definition to make their 
meaning clear. 

\begin{itemize}
\item Consider the Self--Avoiding--Walk problem. The mean square 
end--to--end distance of $N$--step walks and the mean square radius 
of gyration of loops scale as $ \langle R_e^2 \rangle_N \sim 
C N^{2 \nu}$ and $ \langle R_g^2 \rangle_N \sim D N^{2 \nu}$, 
respectively ($N$ is the number of monomers) (Figure 1). The 
critical exponent is fixed by CFT (in two dimensions $\nu = 
\frac{3}{4}$) whereas $C$ and $D$ are amplitudes. By themselves, 
they are not universal but their ratio $C/D$ is free of all scales 
and therefore universal. This number has been theoretically 
predicted by QFT methods in Ref.\,\cite{CMpol}: its value, 
$C/D= 13.70$ is in excellent agreement with its numerical 
determination by series expansions on different lattices 
\cite{Guttmann}. Note that this ratio depends of course 
on the class of universality: for the different class of
universality represented by the Random Walk problem, the above
ratio assumes in fact the value $C/D = 12$.   

\begin{minipage}[b]{.46\linewidth}
\framebox{
\centering\psfig{figure=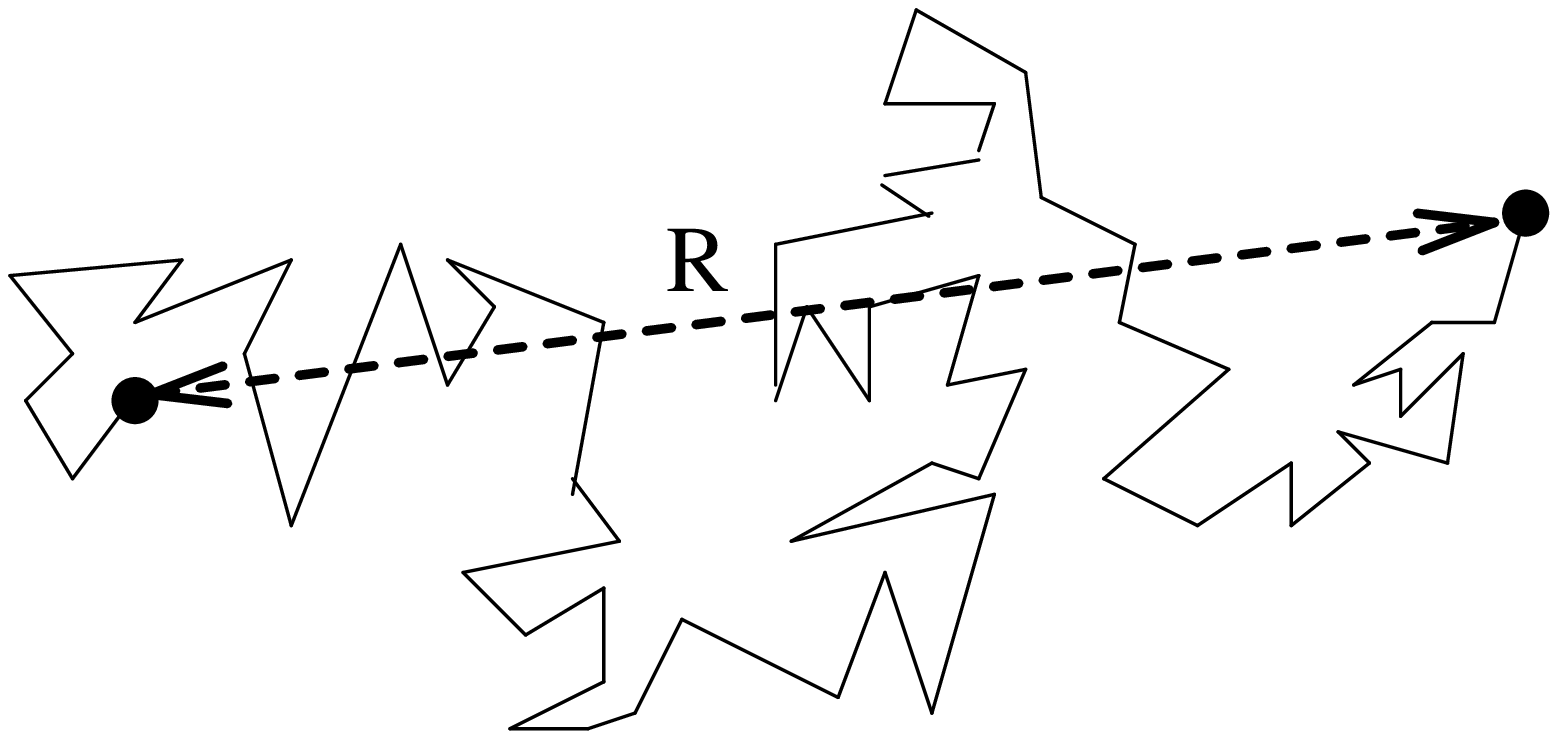,width=\linewidth}
}
\begin{center}
{Figure 1.a: Open $N$--step configuration.}
\end{center}
\end{minipage}  \hfill 
\begin{minipage}[b]{.46\linewidth}
\framebox{
\centering\psfig{figure=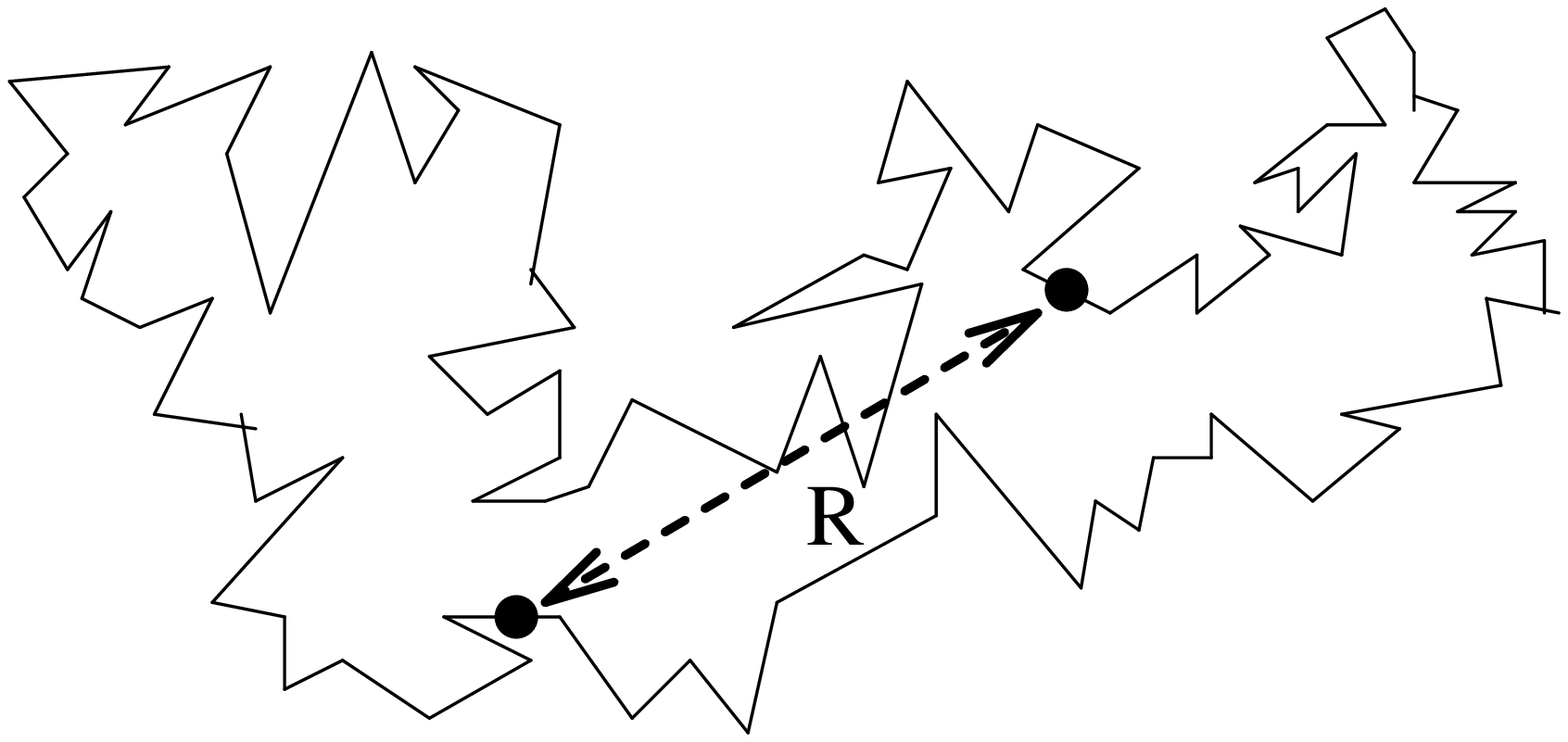,width=\linewidth}
}
\begin{center}
{Figure 1.b: $N$--step loop configuration.}
\end{center}
\end{minipage}

\item Consider the large distance behaviour of the two--point correlation 
function of some order parameter $\Phi$, 
\begin{equation}
\langle \Phi(x) \Phi(0) \rangle 
\sim A_1 
\frac{e^{-m_1 r}}{\sqrt{m_1 r}} + A_2 
\frac{e^{-m_2 r}}{\sqrt{m_2 r}} + A_3 
\frac{e^{-m_3 r}}{\sqrt{m_3 r}} 
\cdots 
\label{expansion}
\end{equation}
where $r = \mid x \mid$. The amplitudes $A_i$ and the mass scales 
$m_i$ entering the exponential falling--off of the correlator are 
not universal: the former depend, in particular, on the normalization 
of the order parameter $\Phi$, the latter on the lattice space and/or 
the value of external fields which spoil the conformal invariance. 
However, their ratios  
\EQ
\displaystyle
\begin{array}{c}
A_2/A_1 \,\,\, , A_3/A_1 \,\,\, \cdots \\
m_2/m_1 \,\,\, , m_3/m_1 \,\,\, \cdots 
\end{array}
\EN 
are universal numbers and, as such, computable by QFT approach. 
For the Ising model in an external magnetic field --- a long--standing 
problem of statistical mechanics ---, the above mass ratios (as well as 
the remaining $m_4/m_1, \ldots, m_8/m_1$ coming from the exact 
solution of the model) have been computed by Zamolodchikov \cite{Zam}, 
with the result $m_2/m_1 = 2 \cos\frac{\pi}{5}=1.618..$; $m_3/m_1 = 2
\cos\frac{\pi}{30}=1.989..$, etc. while the amplitude ratios and 
the scaling form of the two--point function of the magnetization 
operator $\sigma(x)$ have been computed in Ref.\,\cite{DMIs} and 
compared successfully with numerical data (see the figure below, 
where the two--point correlation function $\langle \sigma(x) 
\sigma(0)\rangle$ is plotted versus lattice space distances. The 
point on the graph represent numerical data while the continuum 
curve is the theoretical estimate). Similar calculations have 
been performed for many other statistical models as well: for the 
Tricritical Ising Model, for instance, the above quantities were 
calculated in \cite{ChMus} and \cite{AMV}, respectively; for the
Yang-Lee model can be found in Refs.\,\cite{YL} and \cite{ZamYL}. 

\vspace{-3cm}
\begin{center}
\begin{minipage}[b]{.65\linewidth}
\centering\psfig{figure=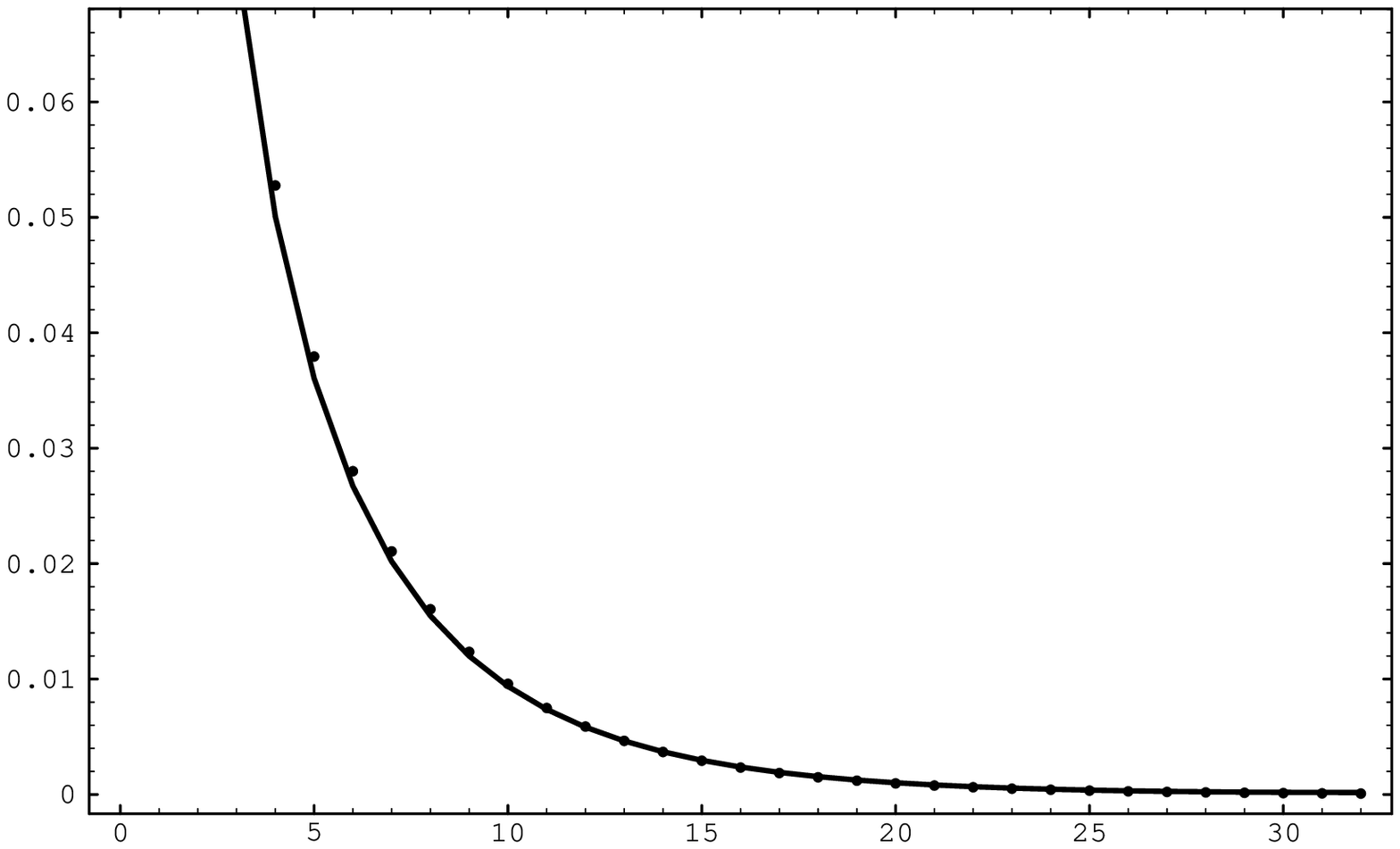,width=\linewidth}

\vspace{-3cm}
\begin{center}
Figure 2, from Ref.\,\cite{DMIs}.
\end{center}
\end{minipage} \hfill 
\end{center}

\item Consider the finite--size effects present in the scaling region of 
a statistical model defined on a cylinder of length $L$ and width $R$, 
with $L/R \gg 1$. Its propagation along the axes $L$ is ruled by the 
Hamiltonian ${\cal H}(R)$. The eigenvalues of this Hamiltonian  --- 
as functions of $R$ --- (Figure 3) turn out to contain quite a large 
number of relevant information. First of all, they have the scaling form  
\EQ
{\cal E}_i(R)=\frac{2\pi}{R}
e_i(m_1 R)\,\,,\hspace{.8cm}i=0,1,2\ldots\,\,\,.
\EN
At very short distance scales, the critical fluctuations are expected
to dominate so that the spectrum must coincide with that of the
conformal point given by \cite{cardy}                                     
\EQ
{\cal E}_i\simeq\frac{2\pi}{R}\left(2
\Delta_i-\frac{c}{12}\right)\,\,,\hspace{.8cm} m_1 R \ll 1\,\, ,
\EN
where $c$ denotes the central charge and $\Delta_i$ the conformal 
dimensions of the scaling fields in the underlying conformal theory. 
In the infrared limit, on the other hand, one should recover the
spectrum of the massive theory and therefore the energy levels
are given by
\EQ
{\cal E}_i\simeq {E}_{vac} R + {\cal M}_i\,\,,\hspace{.8cm} m_1 R
\gg 1   
\label{spectrum}
\EN
where the first term takes into account the vacuum bulk energy
contribution and ${\cal M}_i$ denotes the mass-gap of the i-th level. 
Therefore, mass ratios and other universal numbers, as for instance 
the combination $E_{vac}/m_1^2$, can be directly extracted from such
kind of numerical data. By considering once again the Ising model in a
magnetic field as an example, it is easy to check that the
determination of the first three mass ratios indeed confirms the 
aforementioned Zamolodchikov's prediction for the spectrum and moreover
that $E_{vac}/m_1^2 = - 1/\left(\sin\frac{\pi}{5} \sin\frac{\pi}{3} 
\sin\frac{\pi}{30}\right) = -0.0617..$ (Figure 3.a) (for a lattice
determination of this universal ratio see \cite{BS}). It should 
be mentioned that a detailed theory of the approaching of the energy levels 
to their asymptotic infinite--volume values can also be developed, and that 
a direct measurement of the elastic scattering $S$--matrix can be 
performed too \cite{Luscher}.

\vspace{-15mm}

\hspace{-25mm}
\begin{minipage}[b]{.56\linewidth}
\centering\psfig{figure=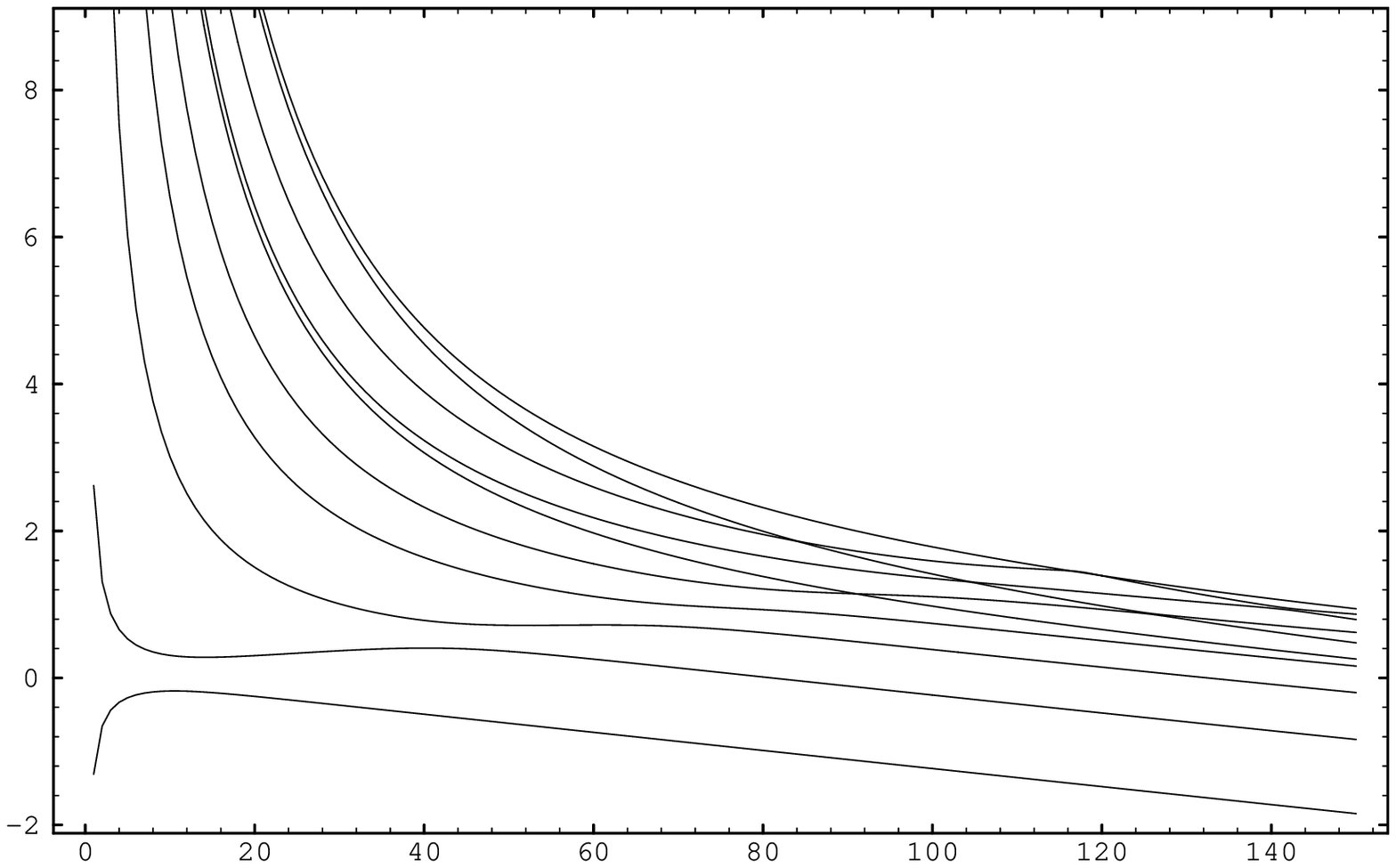,width=\linewidth}

\vspace{-3cm}
\begin{center}
{Figure 3.a: First energy levels of the Ising model at $T=T_c$ 
in a magnetic field versus the width $R$ of the cylider: integrable case.}
\end{center} \end{minipage} \hfill 
\begin{minipage}[b]{.56\linewidth}
\centering\psfig{figure=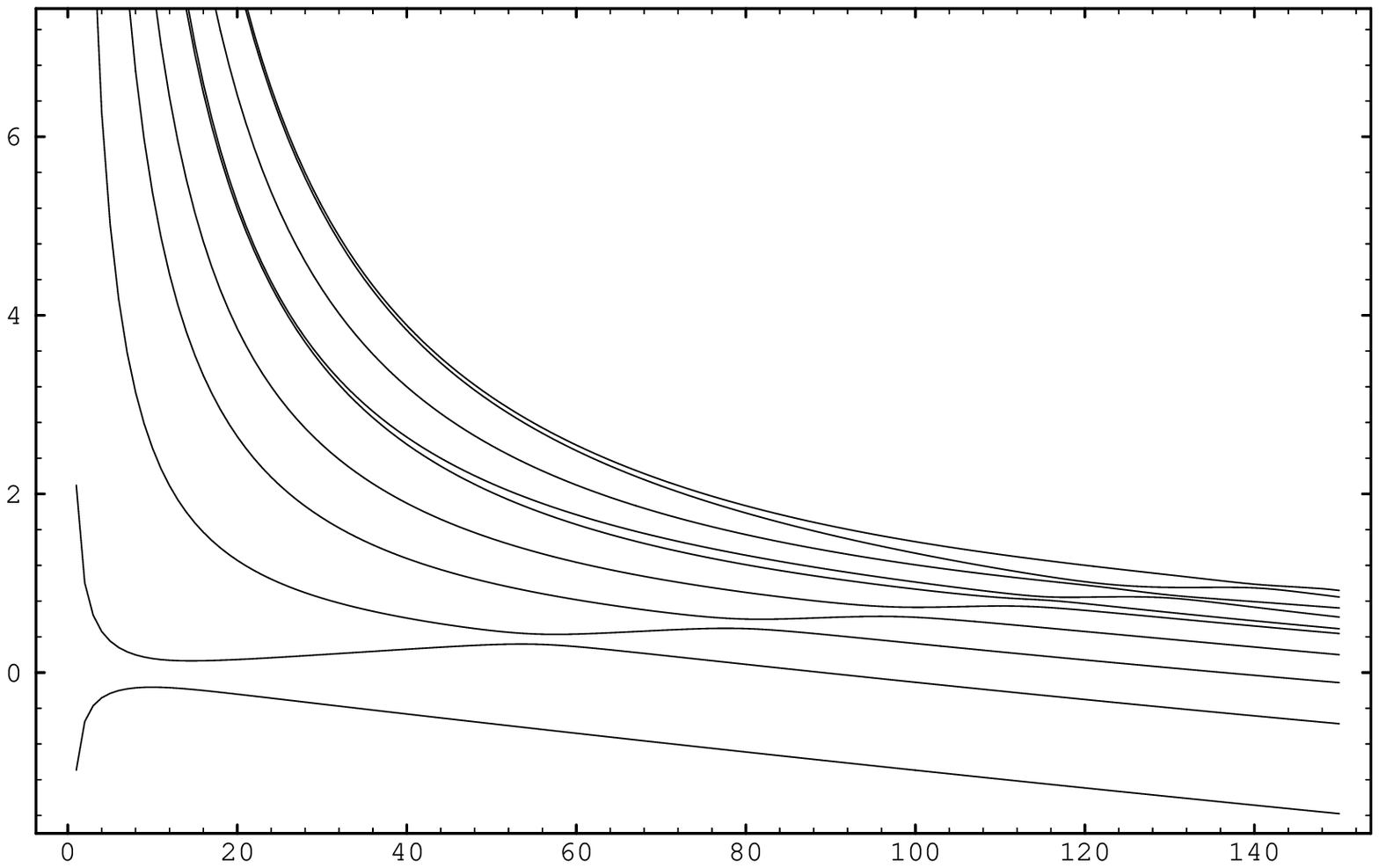,width=\linewidth}

\vspace{-3cm}
\begin{center}
{Figure 3.b: First energy levels of the Ising model at 
$T\neq T_c$ in a magnetic field versus the width $R$ of the cylinder:
non--integrable case.} \end{center} \end{minipage}

\vspace{8mm}

Finally, a few words on level crossings: they are expected to occur 
when the off--critical dynamics is ruled by integrable QFT. On the
contrary, phenomena of level repulsions of the energy lines 
are expected to be observed when we are in presence of non--integrable
theories. This seems to be indeed the case, as illustrated for instance 
by the comparison of Figure 3.a with the Figure 3.b, the latter 
showing the first energy levels of the Ising model, deformed by both 
thermal and magnetic deformations. This double deformation of
the Ising model defines a non--integrable QFT which has been 
extensively analysed in \cite{DMSnonint} and \cite{McCoy-Wu}. 

\end{itemize} 
After a certain familiarity with the topics of interest in this 
paper has been hopefully achieved with the help of the above examples, 
let us now address the main question: how can the scaling region of the 
statistical models be efficiently controlled? The key idea is to consider 
such models as {\em deformations} of the corresponding CFT \cite{Zam}, 
in such a way that their action can be written as
\EQ
{\cal A} \,= \,{\cal A}_{CFT} + \sum_i 
\lambda_i \int \varphi_i(x) \,d^2 x
\,\,\,. \label{action}
\EN  
$\varphi_i$ are scaling fields of conformal dimension $\Delta_i$ 
and, correspondingly, $\lambda_i$ are dimensionful coupling constants 
$\lambda_i \sim \xi^{2 (\Delta_i -1)}$. Within this setting, Conformal 
Field Theory provides a complete description of the short--distance
(ultra--violet) properties and the problem is then to extract the
large--distance behaviour of the field theories (\ref{action}), i.e. 
to study their infrared region. In this respect, of all the possible
deformations the so--called {\em integrable} ones play a very special 
role\footnote{Some criteria can be given to select the integrable
deformations. For a detailed discussion on this point, see 
\cite{Zam,GMrep}.}: they possess an infinite number of conserved 
quantities which gives the possibility of solving them along 
non--perturbative methods. First of all, their on--shell 
characterisation is rather simple because, in virtue of the 
infinite number of conserved charges, all their scattering 
processes are completly elastic and factorizable\footnote{In the 
following we will only refer to massive theories but massless 
integrable theories can be studied as well, by means of a suitable 
extension of the formalism, see \cite{Zammass,DMSml}.} \cite{Zam,ZZ}. 
The whole set of scattering amplitudes (and consequently, the exact
spectrum of the excitations) can be very often explicitly computed 
as solutions of managable functional equations, expressing general
conditions, as unitarity, crossing symmetry and the bootstrap 
equivalence of all excitations, i.e. the absence of distinction 
between bound states and asymptotic particles (for a review, see 
Ref.\,\cite{GMrep}). 

One of the simplest examples of such bootstrap theories is provided by the 
scattering theory of the off--critical Yang--Lee model \cite{YL}: the 
massive theory consists in this case of a single massive excitation $A$, 
which may be regarded as bound state of itself. The exact two--body
$S$--matrix of the model  satisfies in this case the functional equations
\begin{eqnarray}
&& S(\beta) S(-\beta) = 1 \,\,\,;\nonumber \\ 
&& S(\beta) = S(i \pi - \beta) \,\,\,;
\label{YLSmatrix} \\
&& S(\beta) = 
S\left(\beta - i \frac{\pi}{3}\right)  
S\left(\beta + i \frac{\pi}{3}\right)  \,\,\,.\nonumber
\end{eqnarray} 
where the standard parameterisation of two-dimensional on mass-shell
momenta in terms of rapidities is adopted: $p^0 = m\cosh\beta_1$,
$p^1 = m\sinh\beta_1$, $\beta\equiv\beta_1 -\beta_2$. The minimal 
solution of the above equations is then given by 
\EQ
S(\beta) \,=\,\frac
{\tanh\left(\beta + \frac{2 \pi i}{3}\right)}
{\tanh\left(\beta - \frac{2 \pi i}{3}\right)} \,\,\,.
\EN 
Other examples of exactly solvable scattering theories can be found 
in \cite{GMrep}.  

However, for the scope of statistical mechanics, the most important feature 
of the two--dimensional integrable models is that all the Form Factors
(FF) can be exactly determined. These are the matrix elements of the
local operators ${\cal O}(x)$ on the asymptotic states of the theory 
\EQ
\begin{array}{l}
{}_{b_1\ldots b_m}F^{\cal O}_{a_1\ldots
a_n}(\beta'_1,\dots,\beta'_m|\beta_1,\ldots,\beta_n) \equiv \\
\,\,\,\,\,\,\,\,\,
{}^{out}_{\,\,\,\,\,\,0}\langle A_{b_1}(\beta'_1)\ldots
A_{b_m}(\beta'_m)|{\cal O}(0)|
A_{a_1}(\beta_1)\ldots A_{a_n}(\beta_n)\rangle_0^{in}\,\,\,.
\end{array}
\label{matrixelements} 
\EN
Their computation can be performed once the exact $S$--matrix, the bound 
state structure and the asymptotic behaviour of the matrix elements 
(this being fixed by CFT, though) are known: the FF possess, in fact, 
branch cut singularities dictated by the unitarity conditions of the 
$S$--matrix and moreover satisfy an infinite number of recursive 
equations originating from kinematical and bound state poles\footnote{
The reader is referred to the original literature for comprehensive 
studies on this subject.} \cite{DMIs,ZamYL,KW,Smirnov}. Interestingly
enough, for many statistical models close solutions of the above 
conditions satisfied by the FF have been given in terms of elegant 
mathematical formulas, consisting of elementary symmetric polynomials and
determinant expressions thereof (see, for instance \cite{ZamYL,ShG,Smirnov}).   

The knowledge of the Form Factors of the theory has two important
consequences. The first is that we can investigate the off-shell 
behaviour of the theory, i.e. we can compute the two-point (as well 
as higher-point) correlation functions of the model by means of the 
spectral representations obtained through the unitarity sum (Figure 4) 
\begin{eqnarray}
& & \langle \Phi(x) \Phi(0)\rangle = 
\sum_{n=0}^{\infty} \int_{\beta_1 > \beta_2
\ldots \rangle \beta_n} \frac{d\beta_1}{2\pi} \cdots \frac{d\beta_n}{2\pi}
\label{spectral} \\
& & \,\,\,\,\,\,|\langle 0|\Phi(0)|A_{a_1}(\beta_1) \cdots
A_{a_n}(\beta_n) \rangle|^2 e^{-|x| \sum_{k=1}^n m_k \cosh\beta_k} \nonumber
\end{eqnarray}

\vspace{8mm}

\begin{center}
\begin{minipage}[b]{.65\linewidth}
\framebox{
\centering\psfig{figure=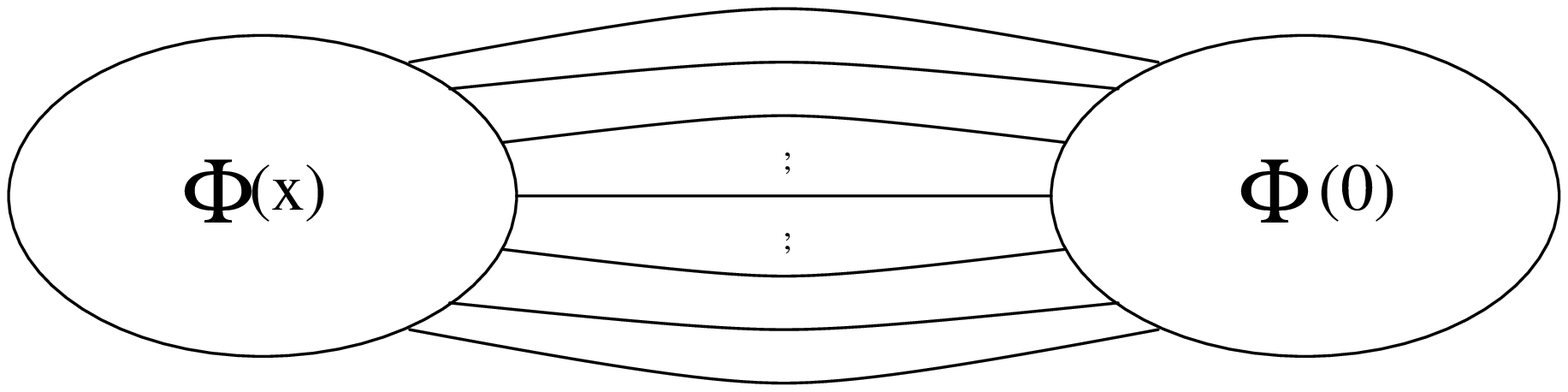,width=\linewidth}
}
\begin{center}
Figure 4.
\end{center}
\end{minipage} \hfill 
\end{center}

A theoretical argument as well as a long practise with spectral
representations of two--dimensional models have shown a quality 
of the foremost practical importance, namely their very fast rate 
of convergence for  {\em all} distance scales (see, for instance 
\cite{CMpol,DMIs,AMV,ZamYL,DMSml}). In view of this property, 
correlation functions of many statistical models have been 
determined with remarkable accuracy by means of limited 
mathematical efforts. 

The second consequence is that the knowledge of the FF permits to 
test non--integrable aspects of statistical models. Non--integrable 
models are of course the rule rather than the exception and many 
statistical systems fall, in fact, in this class of models: the Ising 
model at a generic point of its phase diagram, for instance, or the 
Ashkin--Teller in presence of additional magnetic couplings. Other
examples of non--integrable systems are provided by spin wave 
propagation in anisotropic magnetic liquids or ultra--short optical 
pulses propagating in resonance degenerate medium \cite{Bullough} 
and --- in field theory context --- by the massive Schwinger model 
\cite{Coleman}. The approach proposed in \cite{DMSnonint,DMSG} 
to study non--integrable models consists of viewing them as 
deformations of the integrable ones. In this way, corrections 
to the masses and to the scattering amplitudes can be computed 
in terms of the FF of the operator(s) which spoils integrability, 
in much the same way as done in perturbation theory in quantum 
mechanics \cite{DMSnonint}. However, the breaking of integrability 
has the largest effect on non--trivial topological sectors of the 
statistical model \cite{DMSG}. This is the case of the kinks of 
the Ising in its low temperature phase or the solitons present 
in those models described by equations of Sine-Gordon type. Phenomena 
of confinement and resonance states are also usually observed as 
a consequence of non--integrable perturbations. 

In conclusion, a thorough understanding of integrable statistical 
models and field theories has been achieved in the last decade, 
with the determination of their exact spectra, correlation 
functions, Renormalization Group flows and cross--over phenomena, 
etc. Some progress has also been recently achieved in the study of 
non--integrable models although a more detailed analysis is still 
needed for these models. The complete characterisation of their 
dynamics may be in fact regarded as one of the most important open 
problems of the subject.  

\vspace{15mm} {\em Acknowledgements}. I would like to thank J.L. Cardy, 
G. Delfino and P. Simonetti for their collaboration on these topics.

\end{document}